\documentclass[runningheads]{llncs}
\usepackage{graphicx}
\usepackage{amsmath}
\usepackage{subcaption}
\usepackage{array}
\usepackage{float}
\usepackage{hyperref}
\hypersetup{
    colorlinks=true,
    linkcolor=black,
    filecolor=black,      
    urlcolor=blue,
    citecolor=black,
}

\begin{document}
\title{Registration of ultrasound volumes based on Euclidean distance transform}
\author{Luca Canalini\inst{1,2}\href{}{} \and
Jan Klein\inst{1} \and
Dorothea Miller\inst{3} \and
Ron Kikinis\inst{1,2,4}}
\institute{Fraunhofer MEVIS, Institute for Digital Medicine, Bremen, Germany \email{luca.canalini@mevis.fraunhofer.de}\\
\and
Medical Imaging Computing, University of Bremen, Germany \and
Department of Neurosurgery, University Hospital Knappschaftskrankenhaus, Bochum, Germany \and
%\email{\{abc,lncs\}@uni-heidelberg.de} \and
Surgical Planning Laboratory, Brigham and Women’s Hospital, Harvard Medical School, Boston, USA}
\maketitle              % typeset the header of the contribution
\begin{abstract}
During neurosurgical operations, surgeons can decide to acquire intraoperative data to better proceed with the removal of a tumor. A valid option is given by ultrasound (US) imaging, which can be easily obtained at subsequent surgical stages, giving therefore multiple updates of the resection cavity. To improve the efficacy of the intraoperative guidance, neurosurgeons may benefit from having a direct correspondence between anatomical structures identified at different US acquisitions. In this context, the commonly available neuronavigation systems already provide registration methods, which however are not enough accurate to overcome the anatomical changes happening during resection. Therefore, our aim with this work is to improve the registration of intraoperative US volumes. In the proposed methodology, first a distance mapping of automatically segmented anatomical structures is computed and then the transformed images are utilized in the registration step. Our solution is tested on a public dataset of 17 cases, where the average landmark registration error between volumes acquired at the beginning and at the end of neurosurgical procedures is reduced from 3.55mm to 1.27mm.
\end{abstract}
\section{Introduction}
Before starting a neurosurgical procedure for tumor removal, preoperative data is usually acquired to better plan the successive resection. The most common option is given by magnetic resonance imaging, which can also be accessed during the ongoing surgical procedure to have a better understanding of the resection. In fact, neuronavigation systems can be used to link an intracranial pin-pointed location to the corresponding position in the preoperative data. However, the resection of the tumor and the related anatomical modifications in the surrounding tissues alter the initial configuration of the brain. As consequence, the anatomical structures will be in another conformation with respect to the one observed in the preplanning data~\cite{brainshift}, which soon becomes unreliable during neurosurgery. To obtain an updated view of the resection cavity, neurosurgeons can collect intraoperative US data during the resection itself~\cite{iUS1,iUS2}. These images can be acquired at different stages of the procedure, for example at the beginning of the surgery, just before opening the dura mater, in order to have an initial estimation of which tissues have to be removed. Moreover, a further acquisition can be done at the end of the resection, to detect possible tumor residual. However, the quality of US images decreases in subsequent acquisitions~\cite{degradation}. Thus, for a better comprehension of the US data obtained at the end of the resection, it would be useful to establish a direct mapping between these images and those acquired at the beginning of the surgery, which have a higher quality. A common solution is provided by neuronavigation systems, which can track the US probe locations and compute a registration between the different acquisitions. However, the generally available systems provide a registration solution which is not enough accurate to model the anatomical deformations happening at subsequent stages. Thus, we propose here an automatic method to improve the registration of US volumes acquired at the beginning and at the end of the surgical operation.

In the context of US-US registration for neurosurgical procedures, some solutions have been already proposed to align volumes acquired before and after resection. For example, the authors in~\cite{firstUS} utilized an intensity-based registration method to improve the visualization of volumetric US images. The authors in~\cite{secondUS} developed a non-rigid registration approach, in which they proposed a methodology to discard non-corresponding regions between subsequent US acquisitions. The same method has been used in~\cite{thirdUS}. In another solution~\cite{forthUS}, the authors aimed to improve the previous algorithm by introducing a symmetric deformation field and an efficient second-order minimization for a better convergence of the method. Then, another method to register pre- and post-resection US volumes was proposed by~\cite{fifthUS}, in which the authors presented a landmark-based registration method. More recently, we provided a segmentation-based method to register US volumes: corresponding structures in US volumes are segmented and then used to guide the registration task~\cite{Segmentation-based}.

We introduce here a solution which in the first step utilizes the segmentation results obtained in our previous work. Furthermore, it subsequently applies a Euclidean distance operator on automatically segmented anatomical structures and then uses the transformed masks to guide the registration task.%I could include some examples
\section{Method}
Our experiments are conducted by using \href{https://www.mevislab.de/}{MeVisLab}, on a computer equipped with an Intel Core i7 and a GeForce GTX 1080 (8GB).
\subsection{Euclidean distance transform}
The first step of our method includes the generation of a distance mapping of automatically segmented brain structures. Regarding the segmentation step, the same methodology has been proposed in our previous solution~\cite{Segmentation-based}, where a more detailed description is also available. The anatomical elements utilized in our method are the main sulci and falx cerebri. In fact, they clearly appear in US acquisitions due to their {\itshape hyperechogenicity} and, moreover, remain visible in subsequent stages, representing valid elements to guide the registration task. To perform the segmentation step, we utilized a convolutional neural network (CNN) model based on the 3D U-Net~\cite{3Dunet}. With respect to the original architecture, the original depth is reduced to two levels, and a dropout with a value of 0.4 is introduced in order to prevent the network from overfitting. For training, we manually segment the main hyperechogenic structures of interest in 17 US volumes acquired before resection from~\cite{RESECT}. A patch size of (30,30,30), padding of (8,8,8) and a batch size of 15 samples have been utilized, and the learning rate has been set to 0.001. The best-trained model was saved according to the highest Jaccard index reached during training and then it was used to segment anatomical structures in volumes acquired in the before- and after-resection stages \cite{RESECT}.

Differently from our previous work, a distance mapping is then applied to the automatically generated masks. Regarding this, we can think of a binary image as composed of two different classes, pixels with 0-value in the background (Bg) and pixels with 1-value in the foreground (Fg)
\begin{equation}
I(x,y,z) = \{Fg,Bg\}
\end{equation}
The distance of each pixel of the foreground from the nearest pixel of the background can be computed. The distance mapping $I_d(x,y,z)$ of the whole image can be expressed as
\begin{footnotesize}
\begin{equation}
\begin{aligned}
I_d=\begin{cases}
0 & I(x,y,z)\in\{Bg\}\\
min(||x - x_0, y-y_0, z-z_0||, \forall I(x_0,y_0,z_0)\in Bg) & I(x,y,z)\in\{Fg\}
\end{cases}
\end{aligned}
\end{equation}
\end{footnotesize}
Different distance metrics $||x,y,z||$ can be used to compute the transformation, and one of the most common is the Euclidean distance which computes the L2 norm
\begin{equation}
||x,y,z|| = \sqrt{x^2+y^2+z^2}
\end{equation}
In the proposed methodology, we applied the Euclidean distance transform on the automatically generated masks.  
\subsection{Registration}
The transformed masks are utilized to guide the registration task, which has been modified with respect to our previous solution. The proposed method is a variational image registration approach based on~\cite{book}, in which the correct registration of two volumes corresponds to the global minimum of a discretized objective function. This function is composed of a distance measure, defining the similarity between the deformed template image and the reference image, and a regularizer, limiting the range of possible transformations in the deformable step. In the proposed solution, we respectively chose the normalized gradient field distance (NGF) measure and the curvature regularizer. Moreover, the choice of the optimal transformation parameters has been conducted by using the quasi-Newton l-BGFS~\cite{BFGS}, due to its speed and memory efficiency.
For the registration of the US volumes acquired before and after resection, a solution able to compensate the complex anatomical modifications happening in the resection should be proposed. Thus, our methodology includes an initial parametric step, followed by a non-parametric one. First, the parametric approach utilizes the information provided by the optical tracking systems as an initial guess and then a rigid transformation is performed. In this stage, to speed the optimization process, the images are registered at a resolution one-level coarser compared to the original one. Secondly, the transformation obtained during the parametric registration is used to initialize the non-parametric step. In this stage, to reduce the chance to reach a local minimum, a multilevel technique is introduced: the images are sequentially registered at three different scales. As output of the registration step, the deformed template image is provided.

\section{Evaluation}
\begin{figure}
\centering
\begin{subfigure}[b]{0.3\linewidth}
\includegraphics[height=1.0\textwidth,width=1.0\textwidth]{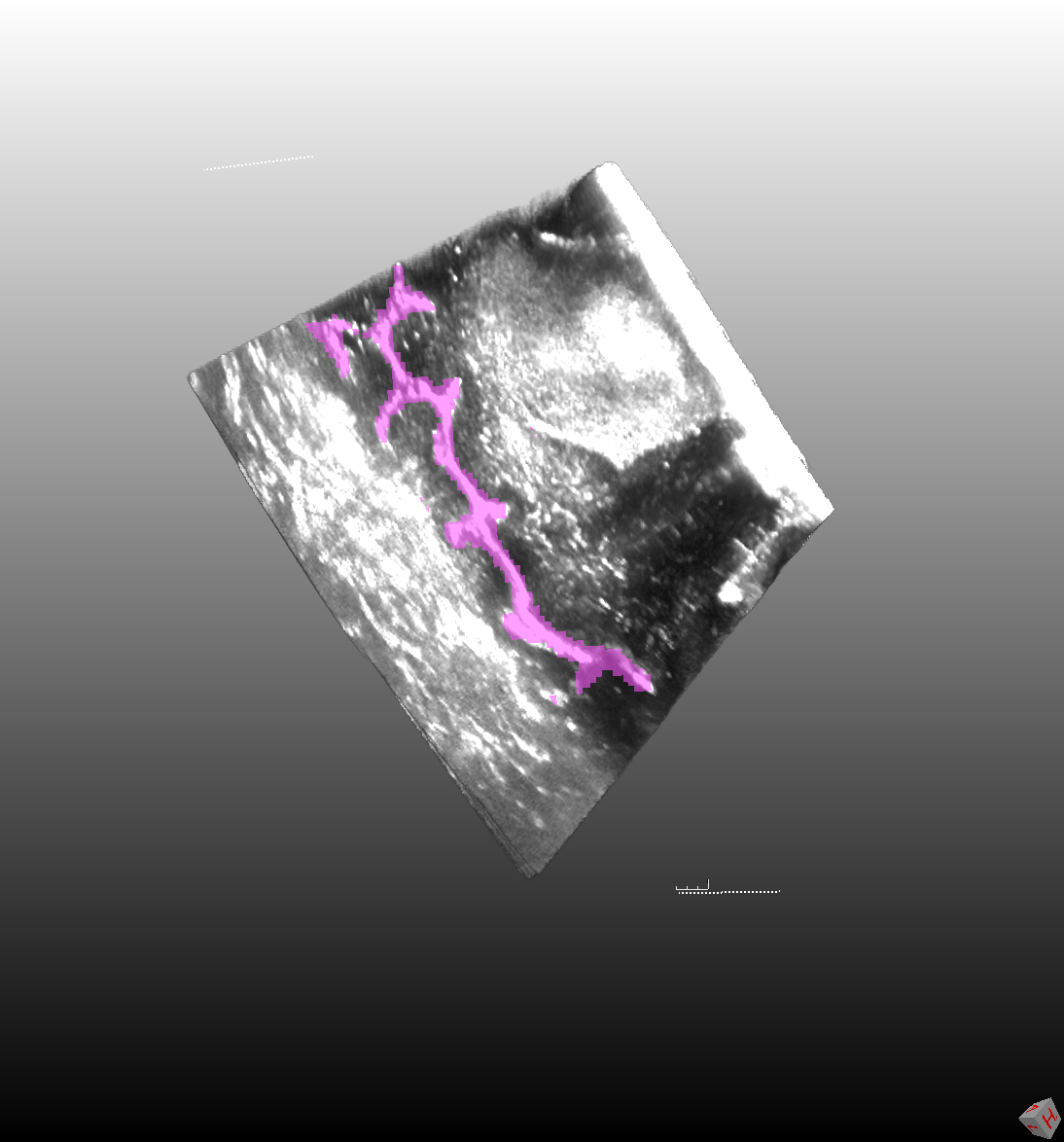}
\caption{}\label{figs2Dbefore}
\end{subfigure}
\begin{subfigure}[b]{0.3\linewidth}
\includegraphics[height=1.0\textwidth,width=1.0\textwidth]{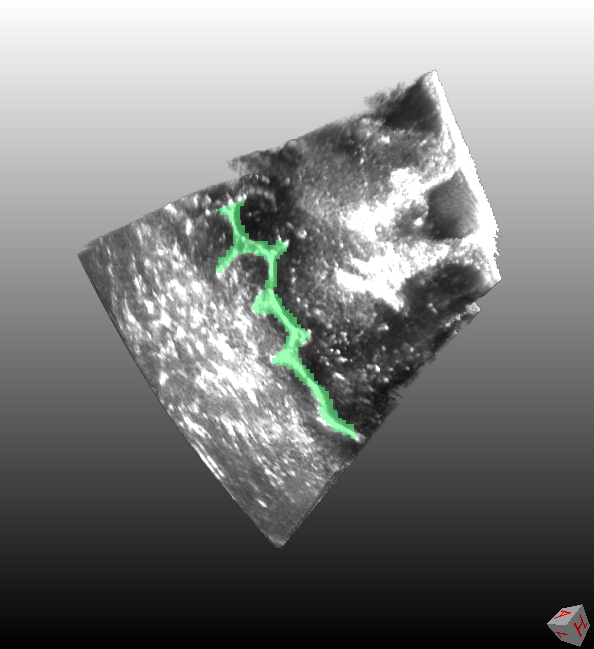}
\caption{}\label{figs2Dafter}
\end{subfigure}
\begin{subfigure}[b]{=0.3\linewidth}
\includegraphics[height=1.0\textwidth,width=1.0\textwidth]{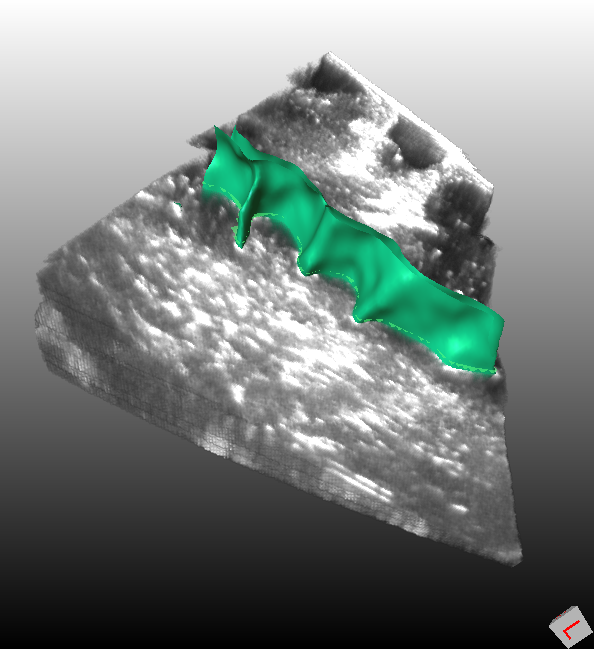}
\caption{}\label{fig3D}
\end{subfigure}
\caption{The same sulcus segmented in corresponding volumes acquired before and after resection stages (Fig.~\ref{figs2Dbefore} and Fig.~\ref{figs2Dafter}). Fig.~\ref{fig3D} shows a partial view of the 3D segmentation of the same structure of Fig.~\ref{figs2Dafter}}
\label{figs2D}
\end{figure}

Our method is tested on 17 cases of the RESECT dataset~\cite{RESECT}. Each case includes two volumes, the first one acquired after craniotomy but before opening the dura mater, the second one at the end of the resection. The corresponding surgical procedures include only resections of low-grade gliomas (tumor of grade II) in adult patients. Corresponding anatomical landmarks are acquired among the two stages and an initial target registration error (TRE) is provided for each patient, together with a mean target registration error (mTRE) and the corresponding standard deviation (sd). In our methodology, the template and reference entries are respectively the volumes acquired before and after resection. The generated deformation field is directly applied to the landmarks acquired after removal, which are therefore registered to the corresponding ones in the pre-section stage.
Regarding the chosen hyperechogenic structures, the first two images of Fig.~\ref{figs2D} show the same sulcus segmented in the volumes acquired before and after resection (Fig.~\ref{figs2Dbefore} and Fig.~\ref{figs2Dafter}). In Fig.~\ref{fig3D} a 3D section of the same structure visualized in Fig.~\ref{figs2Dafter} is provided.
Regarding the registration step, TREs computed before and after applying our registration are available in Table~\ref{tab:table}.
\begin{table}
\centering
\caption{Registration results in millimeter.}\label{tab:table}
\begin{tabular}{|c|c|c|c|}
\hline
\bfseries{Volume} & \bfseries{Landmarks} & \bfseries{Before registration} & \bfseries{After registration}\\
\hline
1 & 13 & 5.80 (3.62 - 7.22) & 1.05 (0.28 - 2.48)\\
2 & 10 & 3.65 (1.71 - 6.72) & 2.32 (0.42 - 4.16)\\
3 & 11 & 2.91 (1.53 - 4.30) & 1.39 (0.55 - 2.24)\\
4 & 12 & 2.22 (1.25 - 2.94) & 0.81 (0.25 - 1.80)\\
6 & 11 & 2.12 (0.75 - 3.82) & 1.62 (0.39 - 4.65)\\
7 & 18 & 3.62 (1.19 - 5.93) & 1.25 (0.25 - 3.15)\\
12 & 11 & 3.97 (2.58 - 6.35) & 0.87 (0.20 - 1.82)\\
14 & 17 & 0.63 (0.17 - 1.76) & 0.62 (0.32 - 1.10)\\
15 & 15 & 1.63 (0.62 - 2.69) & 0.80 (0.27 - 1.81)\\
16 & 17 & 3.13 (0.82 - 5.41) & 1.26 (0.22 - 3.91)\\
17 & 11 & 5.71 (4.25 - 8.03) & 1.51 (0.47 - 5.59)\\
18 & 13 & 5.29 (2.94 - 9.26) & 1.53 (0.30 - 3.61))\\
19 & 13 & 2.05 (0.43 - 3.24) & 1.60 (0.39 - 3.45))\\
21 & 9 & 3.35 (2.34 - 5.64) & 1.82 (0.25 - 5.12)\\
24 & 14 & 2.61 (1.96 - 3.41) & 0.90 (0.24 - 2.33)\\
25 & 12 & 7.61 (6.40 - 10.25) & 1.00 (0.30 - 2.44)\\
27 & 12 & 3.98 (3.09 - 4.82) & 1.24 (0.35 - 2.74))\\
\hline
Mean$\pm$sd & 12.9$\pm$2.6 & 3.55$\pm$1.76 & 1.27$\pm$0.44\\
\hline
\end{tabular}
\end{table}
\begin{figure}
\centering
\begin{subfigure}[b]{0.35\linewidth}
\includegraphics[height=1.0\textwidth,width=1.0\textwidth]{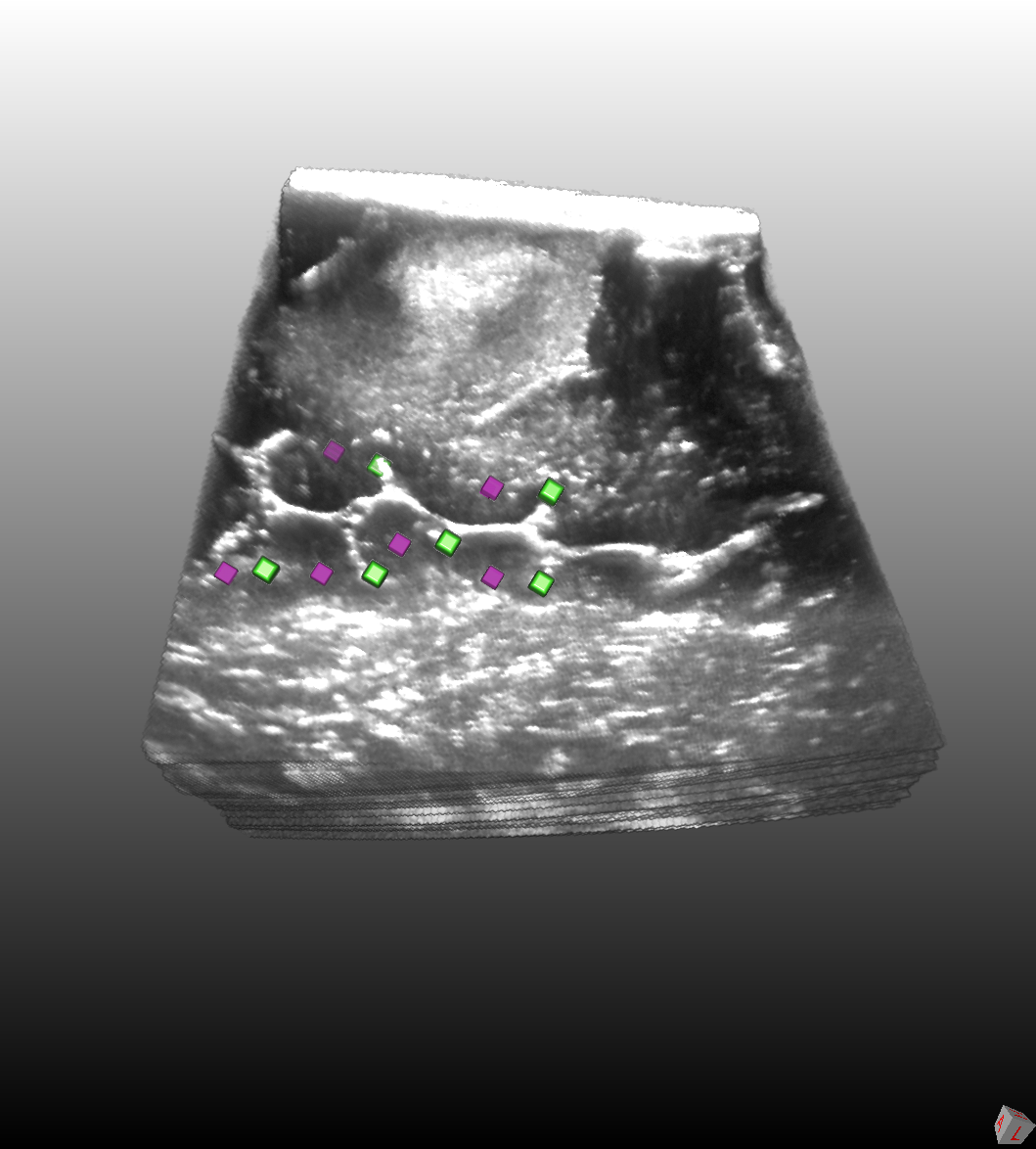}
\caption{}\label{BeforeLand}
\end{subfigure}
\begin{subfigure}[b]{0.35\linewidth}
\includegraphics[height=1.0\textwidth,width=1.0\textwidth]{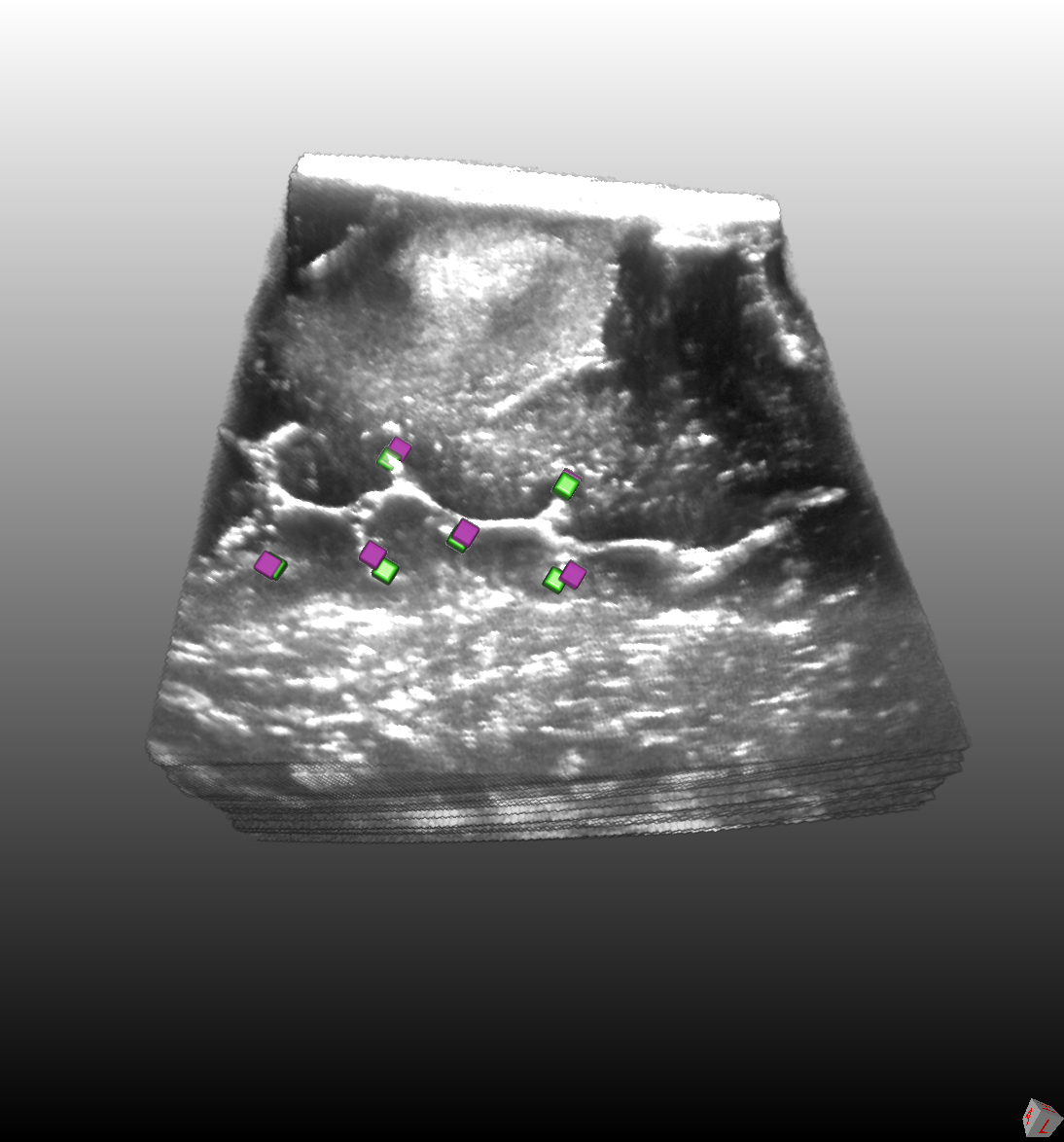}
\caption{}\label{AfterLand}
\end{subfigure}
\caption{Registration results for landmarks. In both images, a 3D section of the volume acquired before resection is provided, with a subset of related landmarks (green). The positions of the landmarks acquired after resection (purple) are provided before and after registration.}\label{landmarks}
\end{figure}
\begin{figure}
\centering
\begin{subfigure}[b]{0.35\linewidth}
\includegraphics[height=1.0\textwidth,width=1.0\textwidth]{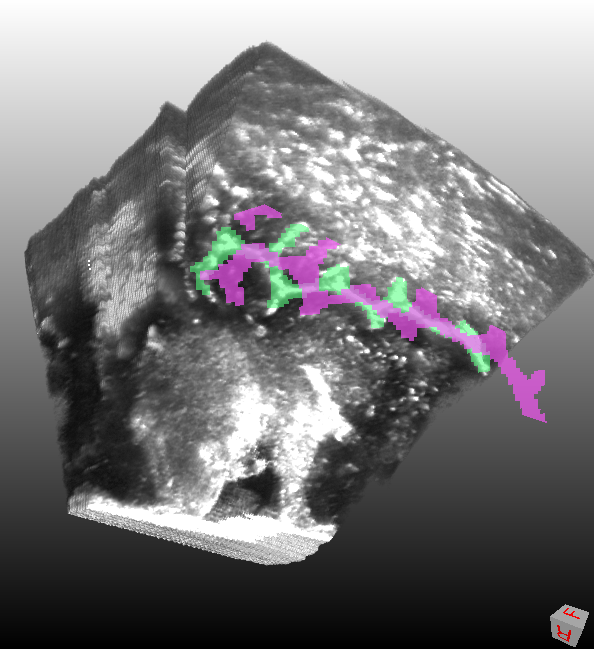}
\caption{}\label{BeforeReg}
\end{subfigure}
\begin{subfigure}[b]{0.35\linewidth}
\includegraphics[height=1.0\textwidth,width=1.0\textwidth]{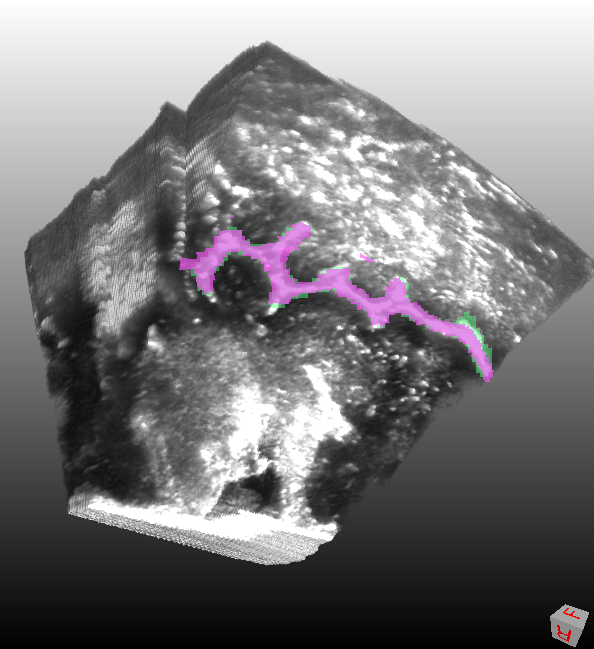}
\caption{}\label{AfterReg}
\end{subfigure}
\begin{subfigure}[b]{0.35\linewidth}
\includegraphics[height=1.0\textwidth,width=1.0\textwidth]{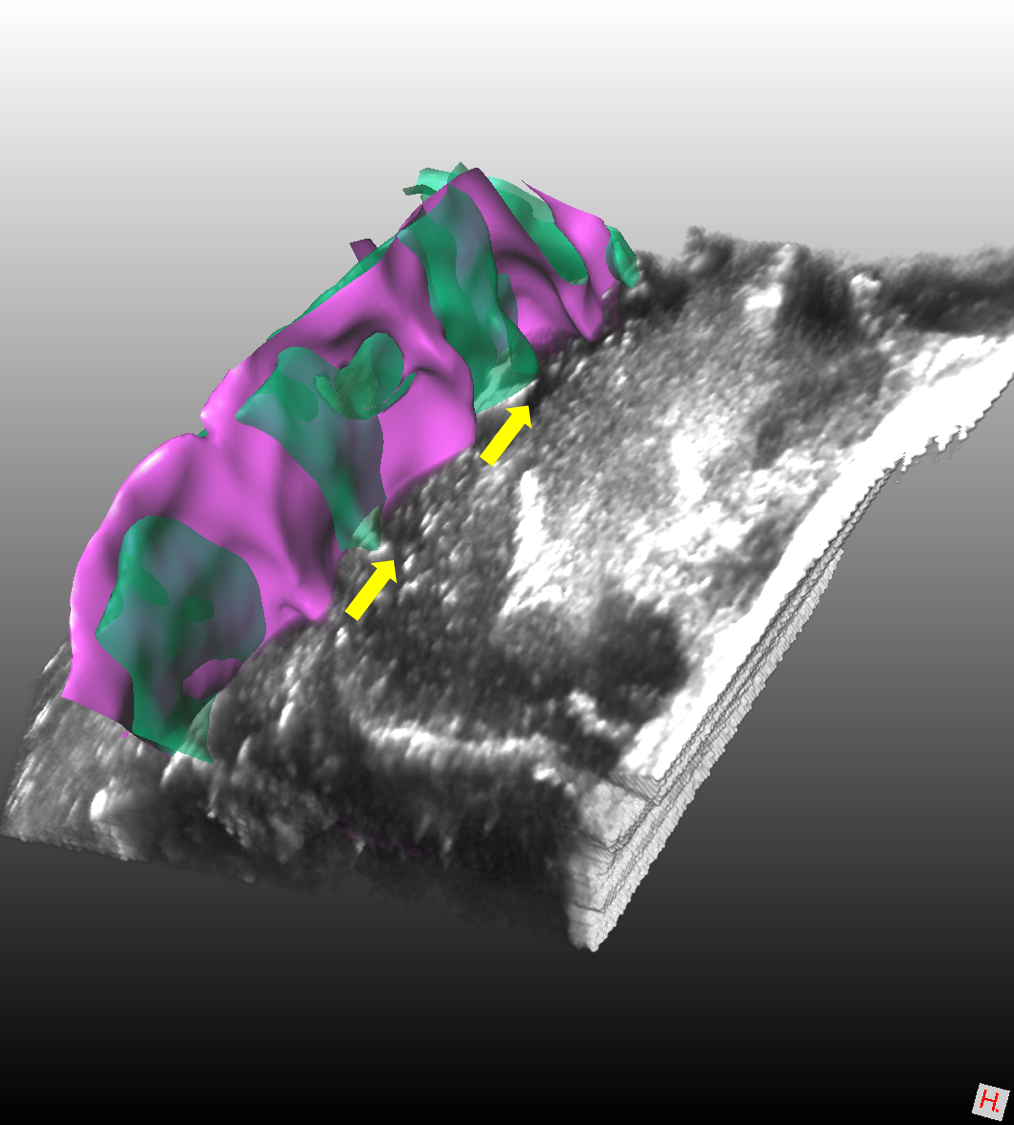}
\caption{}\label{BeforeReg3D}
\end{subfigure}
\begin{subfigure}[b]{0.35\linewidth}
\includegraphics[height=1.0\textwidth,width=1.0\textwidth]{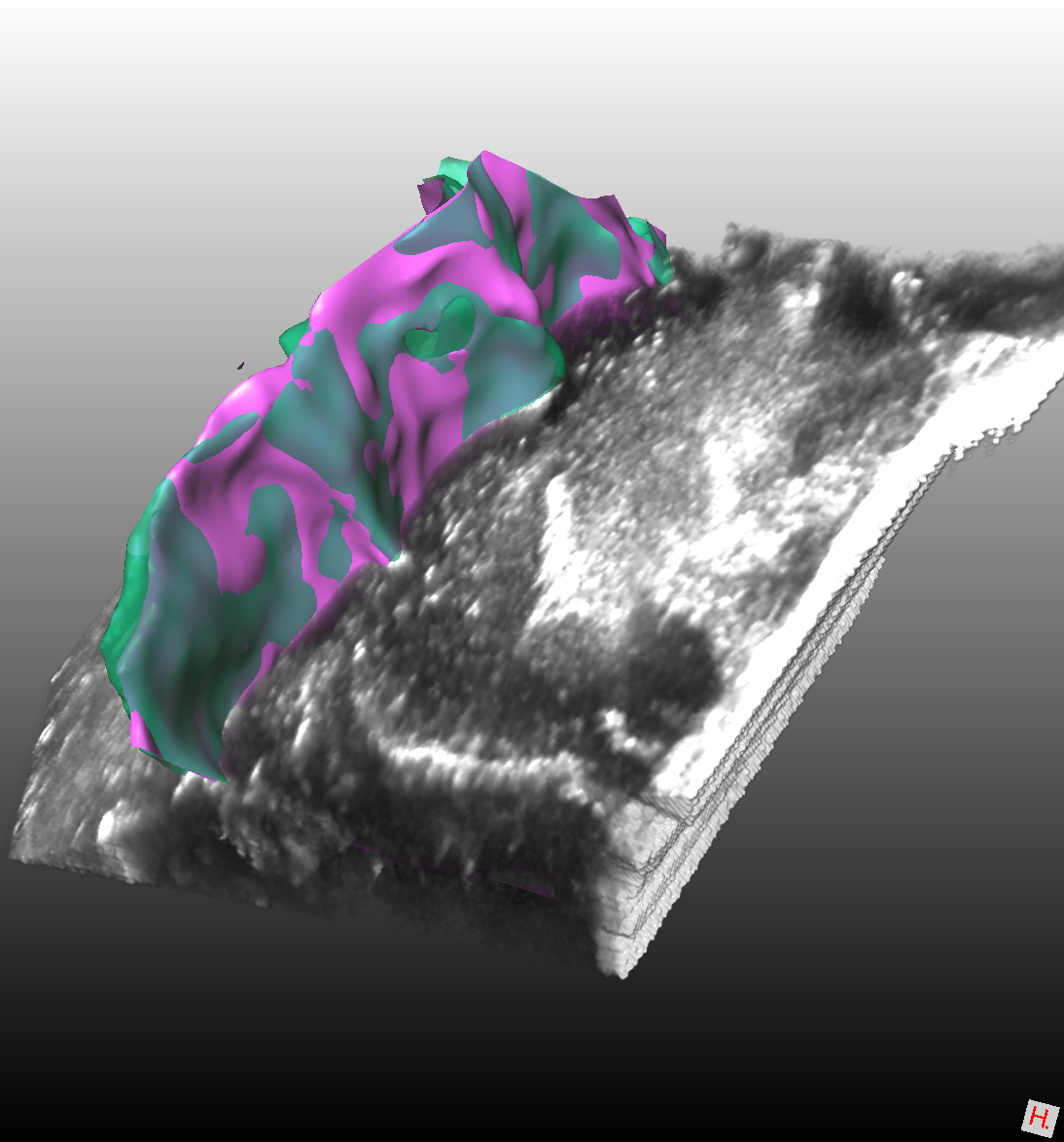}
\caption{}\label{AfterReg3D}
\end{subfigure}
\caption{Registration results for the same sulcus segmented in the before resection (purple) and in the after resection (green) stages. In the first row, a section of the volume acquired after section is displayed, together with 2D views of the segmented structure from both stages. Fig.~\ref{BeforeReg} shows how extended is the original displacement of the masks before registration, which is reduced after applying the proposed method (Fig.~\ref{AfterReg}. In the second row, the same evidence is provided with 3D visualization of the same structure.}\label{FigRegistration}
\end{figure}
By taking as example the same structure of Fig.~\ref{figs2D}, Fig.~\ref{AfterLand} shows the registered landmarks in comparison to the original disposition in Fig.~\ref{BeforeLand}. Moreover, in Fig.~\ref{FigRegistration} the first row displays a section of the volume obtained after resection. Furthermore, Fig.~\ref{BeforeReg} displays the initial displacement between the segmented structure in the pre- and post-resection stages. On the contrary, Fig.~\ref{AfterReg} shows a better overlay between the segmented elements registered with our methodology. In the second row, a section of the same structures is visualized in 3D. Yellow arrows in Fig.~\ref{BeforeReg3D} indicate the correct direction in which the template image should move (Fig.~\ref{AfterReg3D}). The whole procedure, including all steps previously described, takes a mean of 38.34 seconds per each volume.
\section{Discussion}
The hyperechogenic structures of interest are correctly identified in both stages, as shown for the segmented sulcus in Fig.~\ref{figs2D}. Moreover, the chosen structures are useful elements to guide the further registration step. In fact, Table~\ref{tab:table} shows that the initial mTRE is reduced from 3.55 mm to 1.27 mm and the TRE of each case decreases. For the dataset of interest, the proposed method gives proof to correctly register US volumes acquired before and after resection. Visual results related to the registration of the structures of interest in Fig.~\ref{FigRegistration} confirm the numerical findings. Moreover, when the deformation field is applied to the landmarks (Fig.~\ref{landmarks}), we can notice how the updated position of the landmarks acquired after resection is closer to the corresponding landmarks acquired in the volume before resection.
\section{Conclusion}
Our method performs well on the volumes of the RESECT dataset acquired before and after resection. The  proposed solution improves the registration results with respect to our previous work~\cite{Segmentation-based}, which however has been tested on a larger number of cases. Therefore, to better verify the efficacy of the solution, as future work we could decide to apply the proposed solution on a larger set of data.

\section{Acknowledgements}
This work was funded by the H2020 Marie-Curie ITN TRABIT (765148) project. Moreover, Prof. Dr. Kikinis is supported by NIH grants P41 EB015902, P41 EB015898, and U24 CA180918.
%\bibliographystyle{splncs04}

\begin{thebibliography}{8}

\bibitem{brainshift}
Gerard, IJ., Kersten-Oertel, M., Petrecca, K., Sirhan, D., Hall, JA., Collins, DL.: Brain shift in neuronavigation of brain tumors: A review. Med Image Anal. \textbf{35}, 403--20 (2017)

\bibitem{iUS1}
Unsgaard, G., Ommedal, S., Muller, T., Gronningsaeter, A., Nagelhus Hernes, TA.: Neuronavigation by intraoperative three-dimensional ultrasound: initial experience during brain tumor resection.. Neurosurgery \textbf{50}(4), 804--12 (2002)

\bibitem{iUS2}
Unsgaard, G., Rygh, OM., Selbekk, T., Müller, TB., Kolstad, F., Lindseth, F., Hernes, TA.: Intra-operative 3D ultrasound in neurosurgery. Acta Neurochir. \textbf{148}(3), 235--53 (2006)

\bibitem{degradation}
Rygh, OM., Selbekk, T., Torp, SH., Lydersen, S., Hernes, TA., Unsgaard, G.: Comparison of navigated 3D ultrasound findings with histopathology in subsequent phases of glioblastoma resection. Acta Neurochir. \textbf{150}, 1033--1042 (2008)

\bibitem{firstUS}
Mercier, L., Araujo, D., Haegelen, C., Del Maestro, RF., Petrecca, K., Collins, DL.: Registering pre- and postresection 3-dimensional ultrasound for improved visualization of residual brain tumor. Ultrasound Med Biol. \textbf{39}(1), 16--29 (2013)

\bibitem{secondUS}
Rivaz, H., Collins, DL.: Near Real-Time Robust Nonrigid Registration of Volumetric Ultrasound Images For Neurosurgery. Ultrasound Med Biol. \textbf{41}(2), 574--87 (2015)

\bibitem{thirdUS}
Rivaz, H., Collins, DL.: Deformable registration of preoperative MR, pre-resection ultrasound, and post-resection ultrasound images of neurosurgery. Int J Comput Assist Radiol Surg. \textbf{10}(7), 1017--28 (2015)

\bibitem{forthUS}
Hang, Z., Rivaz, H.: Registration of pre- and postresection ultrasound volumes with
noncorresponding regions in neurosurgery. IEEE Journal of Biomedical and Health
Informatics. \textbf{20}, 1240--1249 (2016)

\bibitem{fifthUS}
Machado, I., Toews, M., Luo, J., Unadkat, P., Essayed, W., George, E., Teodoro, P., Carvalho, H., Martins, J., Golland, P., Pieper, S., Frisken, S., Golby, A., Wells, W. 3rd: Non-rigid registration of 3D ultrasound for neurosurgery using automatic feature detection and matching. Int J Comput Assist Radiol Surg. \textbf{13}(10), 1525--1538 (2018)

\bibitem{Segmentation-based}
Canalini, L, Klein, J., Miller, D., Kikinis, R.: Segmentation-based registration of ultrasound volumes for glioma resection in image-guided neurosurgery. Int J Comput Assist Radiol Surg. \textbf{Online first}, (2019)

\bibitem{3Dunet}
{\c{C}}i{\c{c}}ek, {\"{O}}., Abdulkadir, A., Lienkamp, S.S., Brox, T., Ronneberger, O.: 3D U-Net: Learning Dense Volumetric Segmentation from Sparse Annotation. CoRR, 1606.06650 (2016)

\bibitem{RESECT}
Xiao, Y., Fortin, M., Unsgård, G., Rivaz, H., Reinertsen, I.: REtroSpective Evaluation of Cerebral Tumors (RESECT): A clinical database of pre‐operative MRI and intra‐operative ultrasound in
low‐grade glioma surgeries. Med Phys \textbf{44}(7), 3875--3882 (2017)

\bibitem{book}
Modersitzki, J.: Flexible algorithms for image registration. SIAM (2009)

\bibitem{BFGS}
Liu, DC., Nocedal, J.: On the limited memory BFGS method for large scale optimization. Mathematical Programming \textbf{45}(1-3), 503–528 (1989)
\end{thebibliography}
%

\end{document}